\documentclass[twocolumn,superscriptaddress,aps,pre]{revtex4-2}

\usepackage{fix-cm}
\usepackage{url}
\usepackage{amsmath,amssymb,amsfonts,latexsym,bm}
\usepackage{graphicx,epsfig}
\usepackage{fancyvrb}
\usepackage{booktabs}
\usepackage{soul}
\usepackage{algorithm}
\usepackage{algpseudocode}
\usepackage{ragged2e}
\usepackage[svgnames]{xcolor}
\usepackage[colorlinks,linkcolor=MediumBlue,citecolor=MediumBlue,urlcolor = MediumBlue]{hyperref}

\allowdisplaybreaks

\makeatletter

\begin{document} 


\title{Smart strategies to navigate turbulent odor plumes reorienting to local wind} 

\author{Lorenzo Piro}\email{lorenzo.piro@roma2.infn.it}
\affiliation{Department of Physics \& INFN, Tor Vergata University of Rome, Via della Ricerca Scientifica 1, 00133 Rome, Italy}
\author{Maurizio Carbone}
\affiliation{Department of Physics \& INFN, Tor Vergata University of Rome, Via della Ricerca Scientifica 1, 00133 Rome, Italy}
\author{Luca Biferale}
\affiliation{Department of Physics \& INFN, Tor Vergata University of Rome, Via della Ricerca Scientifica 1, 00133 Rome, Italy}
\author{Massimo Cencini}
\affiliation{Istituto dei Sistemi Complessi, CNR, Via dei Taurini 19, 00185 Rome, Italy}
\affiliation{INFN ``Tor Vergata", Via della Ricerca Scientifica 1, 00133 Rome, Italy}
\author{Robin A. Heinonen}
\affiliation{Machine Learning Genoa Center (MaLGa) \& Department of Civil,
Chemical and Environmental Engineering, University of Genoa, Genoa, Italy}
\author{Marco Rando}
\affiliation{Universit\'e C\^ote d'Azur, Inria, CNRS, LJAD, Nice, France}
\author{Agnese Seminara}
\affiliation{Machine Learning Genoa Center (MaLGa) \& Department of Civil,
Chemical and Environmental Engineering, University of Genoa, Genoa, Italy}

\date{\today}

\begin{abstract}
    Olfactory search in turbulent environments is a sensorimotor problem that many animals solve with remarkable efficiency, yet replicating this ability in artificial systems is an enduring challenge because detections are intermittent and wind direction fluctuates strongly, rendering standard search strategies unreliable. We introduce a wind-relative reinforcement-learning framework in which an agent navigates a turbulent plume with a single internal variable --- the elapsed time since the last odor detection --- and selects actions relative to a locally estimated wind direction filtered through an exponential memory kernel. Policies are trained and evaluated in direct numerical simulations of turbulence, capturing the multi-scale characteristics of velocity and odor fields in natural environments, both in the presence and absence of a mean wind. 
    In a mild mean wind, the learned policy outperforms well-known biomimetic policies such as cast-and-surge regardless of the wind memory time, yet adapts its movement pattern to wind-estimation quality. In isotropic turbulence, performance peaks at an intermediate wind memory time, identifying temporal wind integration as a regime-dependent resource.
    Our results highlight the importance of developing and validating olfactory-navigation strategies under realistic turbulent conditions, and offer a compact design principle for minimal robotic olfactory navigation and testable predictions for biological search behavior.
\end{abstract}

\maketitle

\section{Introduction}

Locating the source of an odor in a turbulent environment is a fundamental navigation problem that arises across biological and engineering contexts. 
Entomologists have long known that, for insects, olfactory navigation is a multimodal behavior that often leverages an attractive odor cue and the direction of the wind transporting the odor~\cite{carde-willis2008, matheson2022}. The odor cue alone is insufficient because turbulence breaks scalar fields into irregular, intermittent filaments, so that an agent moving through a plume experiences detections as sparse, unpredictable events rather than as a continuous signal~\cite{biferale1995,shraiman2000,crimaldi2001,celani2014}. Concentration gradients are therefore unreliable as directional cues, making strategies like chemotaxis~\cite{berg1993} ineffective, and must be coupled to wind. Yet, the local wind direction itself is not a straightforward guide, as it fluctuates on timescales shorter than those of the search~\cite{murlis1992}.

These challenges have motivated sustained interest across biology, robotics, and fluid mechanics. On the one hand, a wide range of animals routinely navigate towards odor sources in open environments with no access to global positional information — and crucially, they largely do so by combining olfactory cues with active sensing of local flow direction~\cite{carde2021,vanbreugel2022,reddy2022_review,stupski2024,houle2026}. 
In flying and walking insects, dedicated mechanosensory circuits encode wind direction and integrate it with odor signals to drive goal-directed upwind movement. In Drosophila, for instance, neurons receive convergent input from separate odor and wind pathways, and their sparse activation is sufficient to orient their movement in a reproducible wind-relative direction~\cite{suver2019,matheson2022}. On the engineering side, the same challenge arises in autonomous robots tasked with chemical leak detection or environmental monitoring~\cite{hutchinson2019,wang2023,mansfield2024,fukui2025}.

Algorithms for olfactory navigation largely make use of the wind direction as a primary directional cue~\cite{baker2018algorithms,marques2002olfaction,celani2024_book} and optimized strategies have been developed under a variety of simplifying assumptions that make the navigation problem tractable: access to an allocentric reference frame, the availability of a strong and stable mean wind that provides a reliable upwind direction, or prior statistical knowledge on the entire structure of the odor plume~\cite{vergassola2007,loisy2022,loisy2023,heinonen2023,heinonen2025,piro2026,rando2026,verano2023}.
Among these assumptions, the knowledge of a reliable mean wind direction is key to many proposed algorithms: most strategies use it either explicitly, as a fixed allocentric reference frame, or implicitly, by training and evaluating agents in strong-flow conditions where wind-direction estimation is accurate~\cite{balkovsky2002,heinonen2023,rando2025}. In reality, agents typically lack this prior information because the wind is weak or even absent, or because they are endowed with limited sensory or computational resources. Recent work has begun to relax these constraints by using richer internal states encoding histories of past odor detections~\cite{zhao2022,singh2023}, yet such approaches remain demanding in terms of memory and computation, restricting their applicability in resource-constrained settings.

\begin{figure*}[t!]
\centering
\includegraphics[width=\textwidth]{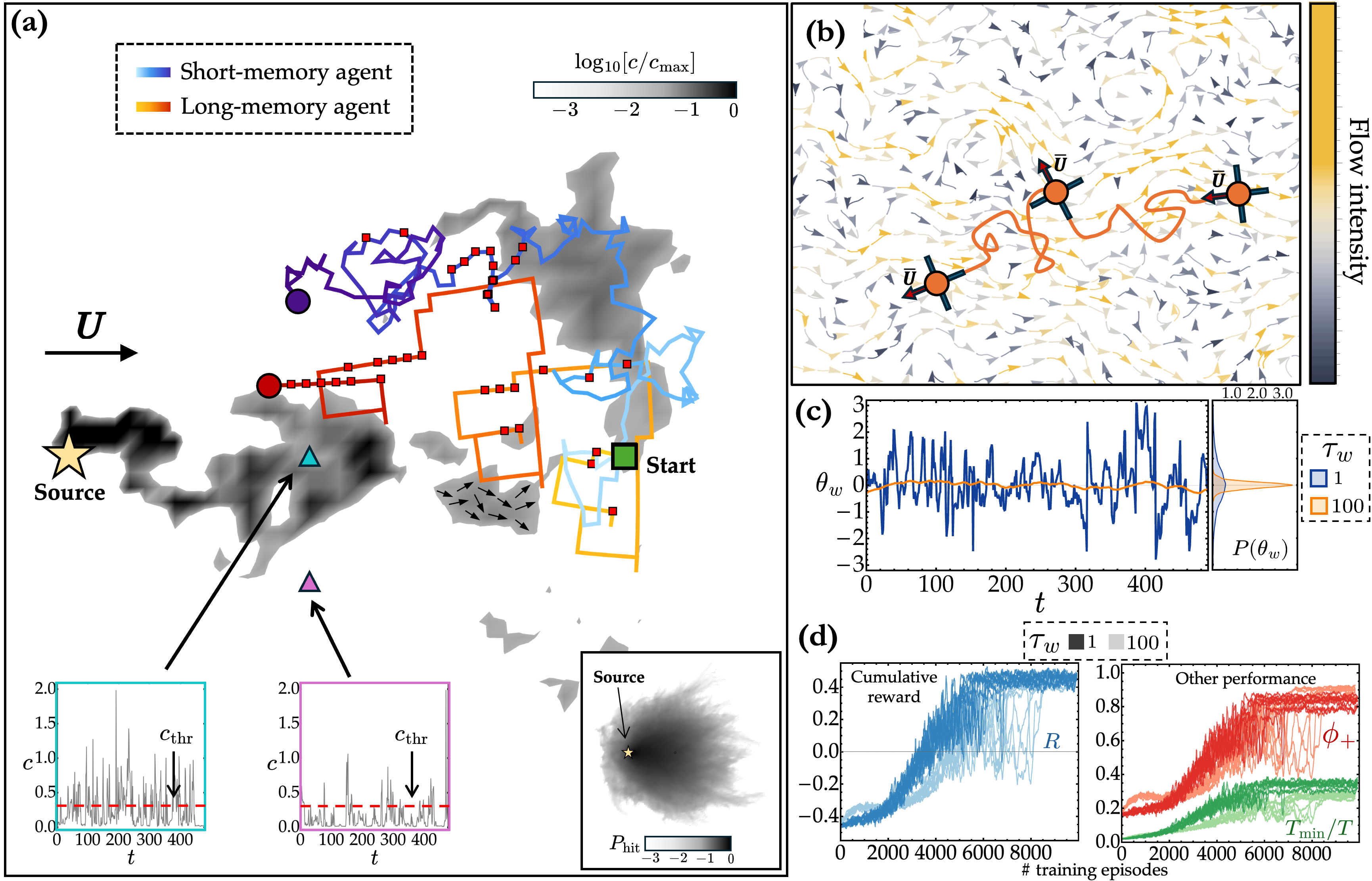}
\caption{\justifying \textbf{Problem setup and wind-relative control framework for navigation in a turbulent odor plume.}
\textbf{(a)} Snapshot of the scalar concentration field (grey scale) for a turbulent plume with mild mean wind $\boldsymbol{U}$ (arrow), with the source marked by a star. Two representative trajectories are shown for agents following learned policies with wind-estimation memory time $\tau_w=1$ (light blue to purple) and $\tau_w=100$ (yellow to red); color gradient along each trajectory encodes time elapsed since departure from the shared starting point (green square), with darker shades indicating later times. Red squares mark odor detection events (hits). 
The black arrows, overlaid on a small portion of the concentration field, illustrate how the local wind direction fluctuates even within a single odor whiff, highlighting the directional uncertainty the agent faces during the search.
Lower insets: concentration time series $c(t)$ recorded at two fixed locations (cyan and magenta triangles), illustrating the intermittent nature of the odor signal and the threshold $c_{\rm thr}$ (dashed red line) used to define a hit. Right inset: empirical detection-probability map $P_{\rm hit}$, showing the spatial probability of encountering the odor plume in this environment. 
\textbf{(b)} Snapshot of the velocity field (arrows colored by local flow intensity) with a schematic of the wind-relative action rule: the agent (orange circle) measures its local wind estimate $\bm{\bar{U}}$ (red arrow) and rotates its discrete action set accordingly.
\textbf{(c)} Estimated wind angle $\theta_w$ time series along the paths shown in panel (a) ($\tau_w=1$, blue; $\tau_w=100$, orange), navigating the same flow realization. Right inset: marginal distributions $P(\theta_w)$ for both memory times, showing that the long-memory estimate is sharply peaked near the mean wind direction while the short-memory estimate is broadly distributed.
\textbf{(d)} Representative training curves for $\tau_w=1$ (dark) and $\tau_w=100$ (light). Left panel: cumulative reward $R$ as a function of training episodes, showing convergence to a stable policy. Right panel: fraction of successful trajectories $\phi_+$ (red) and inverse of the normalized arrival time $T_{\rm min}/T$ (green) over training, confirming that both reliability and efficiency improve and saturate.}
\label{fig:1}
\end{figure*}

Here, we introduce a minimal reinforcement-learning~\cite{sutton2018} framework in which agents learn to navigate turbulent plumes with no prior knowledge of either wind or odor statistics.  
The agent operates solely on local sensory information and a minimal memory structure, with no access to allocentric coordinates.
It defines its own reference frame on the fly, orienting a discrete set of actions relative to its current estimate of the local wind direction obtained by filtering instantaneous local velocity measurements. The agent's internal state is the elapsed time since the last odor detection, which captures the temporal structure of plume intermittency while remaining compact enough to support exact tabular learning; see~\cite{rando2026} for further details on this minimalist memory of odor detections.

Policies are trained using tabular Q-learning~\cite{sutton2018} in two-dimensional direct numerical simulations (DNS) of the Navier-Stokes equations coupled to passive scalar transport sustained by a point-like source, with the objective of finding the source quickly and reliably.

Strategies are learned and evaluated across two complementary flow regimes: a turbulent plume with mild mean wind, where flow fluctuations are strong but a statistically reliable upwind direction exists, and the isotropic limit, where no preferred large-scale direction is available.
We contrast the two regimes and find that directional information can be efficiently encoded by dynamically aligning with the estimated wind direction.
Performance is benchmarked against dedicated heuristic baselines: the biologically inspired heuristic \emph{cast-and-surge} policy~\cite{balkovsky2002}, in which the agent surges upwind upon a detection and casts laterally across the estimated wind direction upon plume loss, in the mean-wind regime, and a spiral-search policy~\cite{masson2009,barbieri2011}, which systematically explores space without any directional bias, in the isotropic limit.


\section{Problem setup and task formulation}

We consider an agent navigating a two-dimensional turbulent odor plume with the objective of locating the odor source as reliably and efficiently as possible within a finite time horizon $T_H$. The environment is generated by direct numerical simulation (DNS) of the two-dimensional Navier--Stokes equations, in the inverse energy cascade regime, coupled to a passive scalar field~\cite{boffetta2012}, originating from a localized source emitting a continuous plume that is stretched and folded by the underlying turbulent flow into an irregular, intermittent structure (see Appendix A for details). We introduce the task in a representative regime with mild mean wind, $U/u_\mathrm{rms} = 1$, where $U$ is the mean wind speed and $u_\mathrm{rms}$ the root-mean-square turbulent velocity fluctuations. This regime serves as the primary setting for explaining both the physical problem and the control framework: a strong mean wind ($U/u_\mathrm{rms} \gg 1$) would render turbulent fluctuations negligible, causing any local wind estimate to converge quickly and reliably to the actual mean flow direction and effectively collapsing the wind-relative frame into an allocentric one (as confirmed by the results shown in Supplementary Fig.~S1). The mild-wind regime, in contrast, places the problem precisely where fluctuations are comparable to the mean flow, making wind-direction estimation a genuine sensorimotor challenge and a meaningful control parameter.

The main difficulty of the task is odor intermittency. Rather than navigating a smooth concentration gradient, the agent's odor encounters are sparse, unpredictable events whose timing and spatial distribution are governed by the statistics of turbulent scalar transport. A detection event, or ``hit'', is registered whenever the local scalar concentration exceeds a prescribed threshold $c_\mathrm{thr}$; this binary criterion constitutes the minimal sensory model adopted throughout.
Figure~\ref{fig:1}(a) shows a representative snapshot of the concentration field together with concentration time series $c(t)$ recorded at two fixed locations within the plume, illustrating the intermittent character of odor exposure. The right inset of Fig.~\ref{fig:1}(a) shows the empirical detection-probability map $P_\mathrm{hit}$, encoding the time-averaged probability of encountering the plume at each location. Together, the time series and the probability map provide a complete statistical picture of the environment that the agent must implicitly learn to exploit by trial and error.

\begin{figure*}[t!]
\centering
\includegraphics[width=\textwidth]{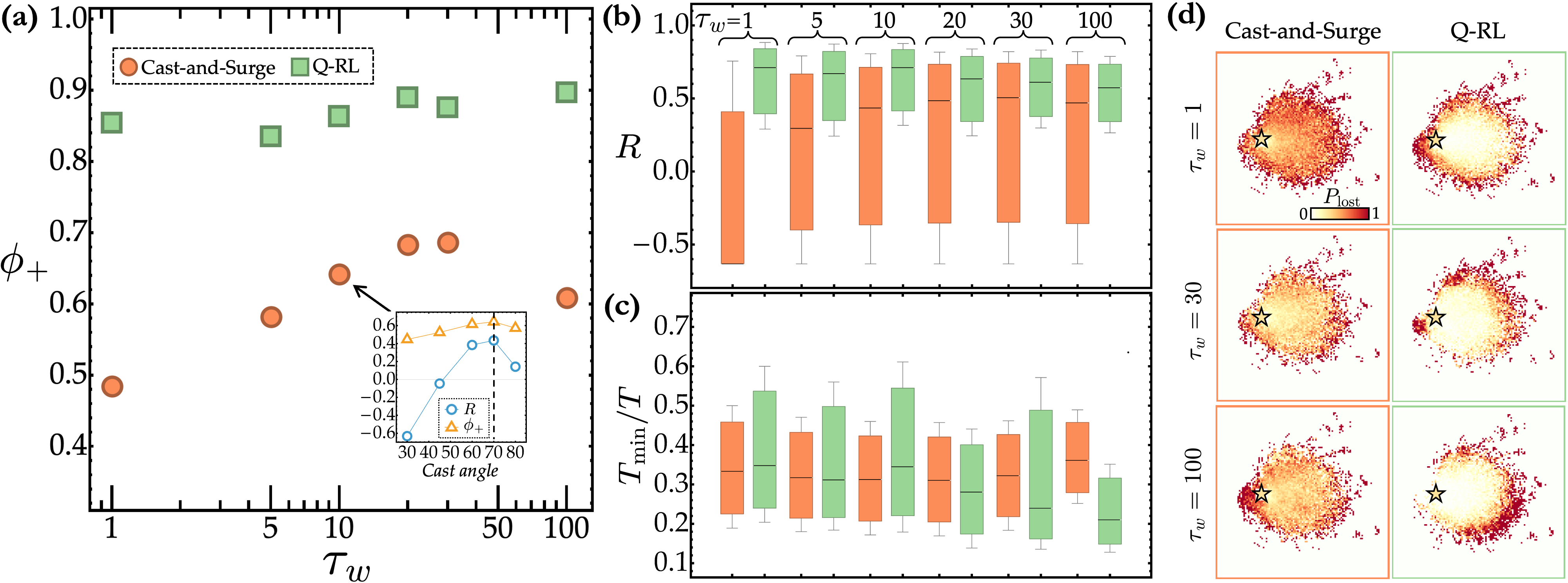}
\caption{\justifying \textbf{Performance of learned policies versus cast-and-surge in the presence of a mild mean wind.}
\textbf{(a)} Success fraction $\phi_+$ (fraction of trajectories reaching the source
within the time horizon) as a function of wind-estimation memory time $\tau_w$ for Q-RL (green squares) and cast-and-surge (orange circles). Q-RL consistently outperforms cast-and-surge across all memory values, while its performance remains nearly independent of $\tau_w$. Inset: tuning of the cast-and-surge casting angle, showing the values of $\phi_+$ and mean reward $R$ used to select the optimal casting angle for the baseline.
\textbf{(b)} Box plots of the cumulative reward $R$ for Q-RL (green) and cast-and-surge (orange) at each tested memory value. Q-RL achieves systematically higher and less variable rewards, reflecting both greater reliability and more efficient trajectories. 
\textbf{(c)} Box plots of the reciprocal of the normalized arrival time $T_{\rm min}/T$ conditioned on success. 
\textbf{(d)} Spatial maps of the failure probability $P_{\rm lost}$ conditioned on the initial position, i.e., the fraction of trajectories starting from each cell that fail to reach the source within $T_H$, shown for cast-and-surge (left column, orange-bordered) and Q-RL (right column, green-bordered) at three representative memory values ($\tau_w = 1$, $30$, $100$, rows from top to bottom). The source location is marked by a star.
Note that the color scale reflects \emph{local} failure rates and should not be used to compare the overall performance of the two strategies, shown in panels~(a-c).
In fact, the more intense red regions visible in the Q-RL maps do not indicate worse overall performance, but rather that the residual failures of Q-RL are spatially concentrated near the intrinsically most challenging initial positions (plume boundary, low-$P_\mathrm{hit}$ regions), whereas cast-and-surge fails more uniformly across the domain. Results shown in all panels are obtained from $5\cdot10^4$ distinct initial conditions.}
\label{fig:2}
\end{figure*}

The agent's internal state is deliberately minimal: a single scalar $\tau_d$, the elapsed time since the last hit, which encodes the temporal structure of plume intermittency without storing any history of past positions, velocities, or detections.
At each discrete time step, the agent selects one of four actions --- upwind, downwind, or either crosswind direction --- whose orientations are rotated to align with its current estimate of the local wind direction $\bar{\bm{U}}$, defining a wind-relative reference frame (Fig.~\ref{fig:1}(b)). This frame of reference mirrors recent results in the fly's fan-shaped body, where the neural circuits that trigger odor-guided navigation relative to the wind were discovered~\cite{matheson2022}. 
In turbulent flows, owing to the roughness of the velocity field~\cite{boffetta2012,frisch1995}, wind direction is a random variable that fluctuates rapidly across nearby points, as visualized in Fig.~\ref{fig:1}(b). The direction of the wind varies even within single whiffs of odor, as these are extended in space (exemplified in Fig.~\ref{fig:1}(a)). The local wind direction $\bar{\bm{U}}$ is estimated by exponentially filtering successive instantaneous local velocity measurements with a characteristic decay time $\tau_w$, the wind memory time, which controls the temporal reach of wind-direction sensing (see Eq.~\eqref{eq_expKernel} in Appendix B). Small $\tau_w$ yields a rapidly reactive estimate that tracks each turbulent fluctuation in real time, while large $\tau_w$ produces a smooth, slowly varying estimate that converges toward the long-run mean wind direction at the cost of a delayed response to local changes. The consequences of this trade-off are made explicit in Fig.~\ref{fig:1}(c), which shows the estimated wind-angle time series $\theta_w(t)$ for the two extreme values of $\tau_w$ in the same flow realization. Short memory produces a broadly distributed, rapidly oscillating signal, while long memory yields a nearly steady estimate, sharply peaked near the true mean wind angle, as confirmed by the marginal distributions $P(\theta_w)$. 
Against this backdrop, Fig.~\ref{fig:1}(a) also shows two representative agent trajectories starting from identical initial conditions and following policies learned with short and long wind memory times. Even at this qualitative level, the two paths differ markedly in geometry, offering a first hint that the wind memory time does not merely tune a filtering parameter but shapes the very structure of the learned search behavior.

To learn these strategies, the compact one-dimensional state space (i.e., the ``clock'' $\tau_d$) is well-suited to exact tabular Q-learning~\cite{sutton2018}. By construction, each episode starts immediately after a detection event, so the agent is initialized in the state $\tau_d=0$ (see also Appendix B).
Policies are trained by maximizing the cumulative reward $R = \sum_{t=0}^{T} \gamma^t \, r_t$, where $T$ is the time at which the source is reached, $\gamma < 1$ is the discount factor, and $r_t$ is the immediate reward, equal to $\gamma - 1$ at each time step $t < T$.
A positive terminal reward $r_T = +1$ is given upon reaching the source before the finite time horizon $T_H$, and no terminal reward is given otherwise.
This choice simultaneously incentivizes reliability by rewarding source-finding episodes and efficiency by penalizing every time step through the negative running reward, so that faster trajectories accumulate less penalty before collecting the terminal bonus.
To maximize this objective, the agent estimates the Q-function $Q(\tau_d, a)$ --- the maximum expected cumulative discounted reward from state $\tau_d$ upon taking action $a$ --- iteratively through episodes of interaction with the environment.
The optimal policy is then the simple readout $\pi^*(\tau_d) = \arg\max_a \, Q(\tau_d, a)$.
Because the state space is one-dimensional and discrete, exact tabular learning converges reliably without function approximation~\cite{sutton2018}, and training requires typically a few thousand episodes to reach a stable policy, as confirmed by the smooth convergence of all performance indicators in Fig.~\ref{fig:1}(d): the success fraction $\phi_+$ (fraction of trajectories reaching the source within the time horizon $T_H$), the reciprocal of the normalized arrival time $T_\mathrm{min}/T$ conditioned on success (with $T_\mathrm{min}$ the minimum possible arrival time from the agent's initial position), and the cumulative reward $R$ itself (see Appendix B for full details of the reinforcement learning algorithm).

\begin{figure*}[ht!]
\centering
\includegraphics[width=\textwidth]{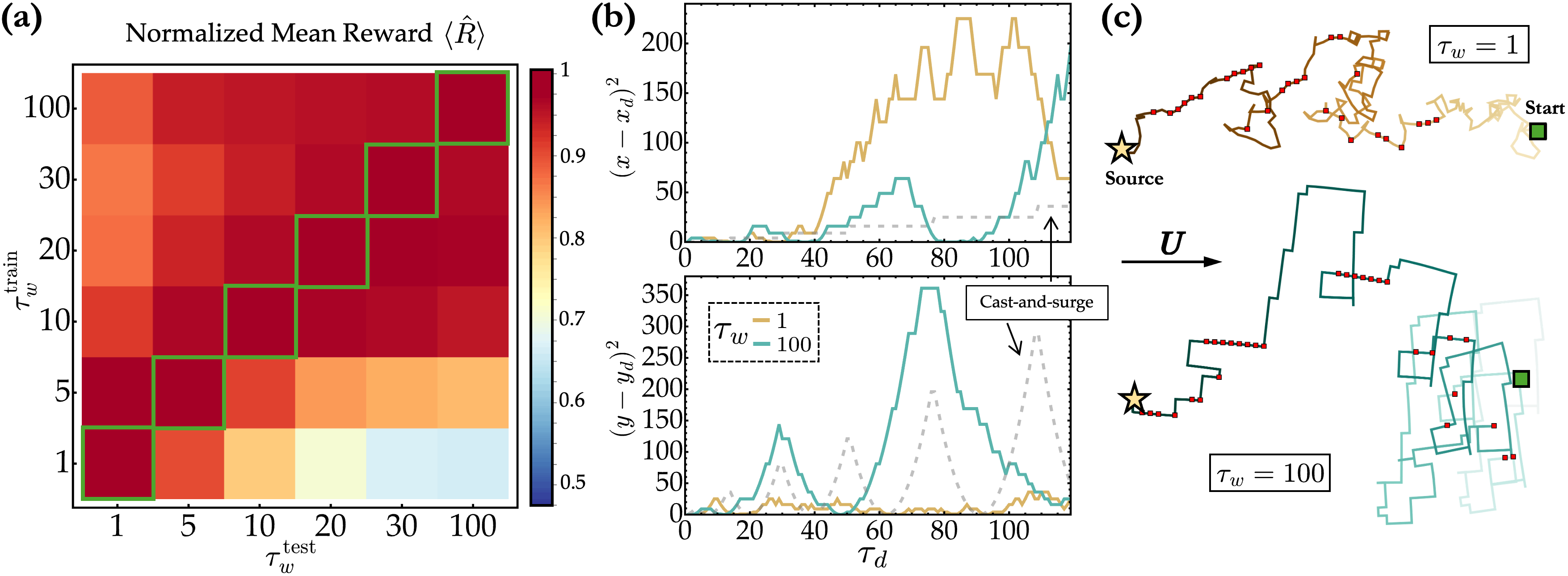}
\caption{\justifying \textbf{Wind memory time governs the geometry of learned
search strategies.} 
\textbf{(a)} Cross-testing matrix showing the normalized mean reward $\langle\hat{R}\rangle$ when a policy trained with wind memory time $\tau_w^\mathrm{train}$ (rows) is deployed with a different operational memory $\tau_w^\mathrm{test}$ (columns), without retraining. Rewards are normalized by the matched diagonal values
(green outline), i.e., the performance of each policy when tested with the same memory used during training. Performance degrades off-diagonal, most strongly at the extremes, yet remains largely preserved for small deviations from the training memory, indicating a smooth transition between strategies across memory regimes.
\textbf{(b)} Conditional mean-squared displacements in the wind-relative reference frame with respect to the last detection position $(x_d, y_d)$: longitudinal component $(x-x_d)^2$ along the estimated wind direction (top) and transverse component $(y-y_d)^2$ (bottom), as functions of $\tau_d$ for $\tau_w = 1$ (gold) and $\tau_w = 100$ (teal). The cast-and-surge baseline is shown for reference (dashed grey). 
\textbf{(c)} Representative trajectories for $\tau_w = 1$ (top, gold tones) and $\tau_w = 100$ (bottom, teal tones), both starting from the same position (green square) with the source marked by a star. Color intensity encodes time elapsed since departure, with darker shades indicating later times; red squares mark detection events.}
\label{fig:3}
\end{figure*}

\section{Learned policies' performance in mild mean wind}

We evaluate navigation performance in the mild-mean-wind regime by measuring the success fraction $\phi^+$, the cumulative reward $R$, and the arrival-time distributions across large ensembles of initial conditions and independent flow realizations, sweeping over a broad range of wind memory times $\tau_w$. The results are summarized in Fig.~\ref{fig:2}. As a benchmark, we use cast-and-surge~\cite{balkovsky2002} with its casting angle tuned to maximize baseline performance (inset of Fig.~\ref{fig:2}(a)). See Appendix C for details.
Learned policies trained by Q-learning reinforcement learning (Q-RL) consistently outperform this competitive baseline across all values of $\tau_w$ tested, both in terms of reliability and efficiency: the success fraction of Q-RL reaches $\phi^+ \approx 0.9$, compared to $0.5\lesssim\phi^+\lesssim0.7$ for the optimally tuned cast-and-surge (Fig.~\ref{fig:2}(a)), and the cumulative reward distributions are systematically shifted towards larger, less fluctuating values (Fig.~\ref{fig:2}(b)).
Notably, the primary advantage of the learned policy lies in its higher success rate rather than in intrinsically faster upwind navigation. In fact, when successful trajectories are considered in isolation, the arrival-time distributions of Q-RL and cast-and-surge are broadly comparable (Fig.~\ref{fig:2}(c)): when cast-and-surge does reach the source, it does so in a time similar to Q-RL. This distinction carries physical meaning: the learned strategy is primarily more \emph{robust} --- better at recovering from plume loss and at avoiding unrecoverable excursions into regions of low detection probability --- without compromising overall navigation efficiency.

To characterize failure modes and understand where each strategy breaks down spatially, we compute failure-probability maps $P_\mathrm{lost}$ conditioned on the initial position for both the learned and cast-and-surge strategies (Fig.~\ref{fig:2}(d)). 
Cast-and-surge fails consistently when the agent is initialized at the plume boundary, independently of $\tau_w$, and also when it is initialized upstream of the source, particularly for large $\tau_w$. These failures indicate that the fixed casting rule is insufficient to re-establish contact with the plume under geometrically unfavorable conditions.
In contrast, failures of Q-RL occur less frequently and are predominantly localized near the plume boundary --- i.e., at the intrinsically most challenging initial positions where $P_\mathrm{hit}$ is low --- rather than being broadly distributed across the domain. Failures in the region upstream of the source are instead markedly less prominent and are further reduced as $\tau_w$ increases. This trend reflects the enhanced directional fidelity of the inferred wind estimate at longer memory times.

Taken together, these results establish the consistent superiority of the learned wind-relative policy in the mild-mean-wind setting and set the stage for a finer mechanistic analysis of how $\tau_w$ shapes the internal structure of the learned strategy. In fact, although aggregate performance is only weakly sensitive to $\tau_w$ over a broad range (Fig.~\ref{fig:2}(a,b)), this does not mean that the learned policy is indifferent to wind memory. Whether the agent genuinely adapts the internal structure of its search behavior to $\tau_w$, or instead learns a single robust policy that happens to work across memory times, is a question we address in the next section.

\section{Memory shapes search strategy}

To directly assess the role of memory in the learned behaviors, we perform cross-testing experiments in which a policy trained with wind memory time $\tau_w^\mathrm{train}$ is deployed with a different operational memory $\tau_w^\mathrm{test}$, without retraining (Fig.~\ref{fig:3}(a)). 
Two complementary observations emerge. Performance degrades measurably off-diagonal --- most strongly between the two extremes $\tau_w = 1$ and $\tau_w = 100$ --- demonstrating that the agent genuinely adapts its strategy to the quality of wind direction estimation available during training. Yet the degradation is gradual: policies tested near their training value retain most of their performance, indicating robustness to moderate deviations in $\tau_w$ and a smooth transition between strategies across memory regimes.

\begin{figure*}[ht!]
\centering
\includegraphics[width=\textwidth]{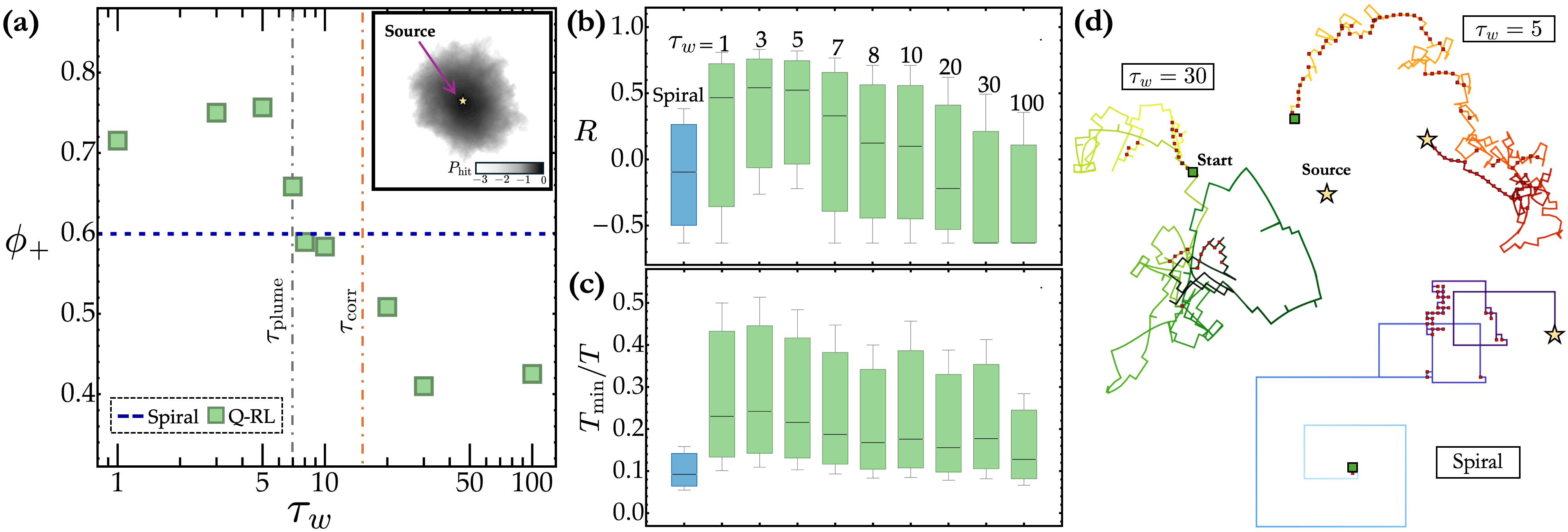}
\caption{\justifying \textbf{Optimal wind-estimation memory in isotropic turbulence.}
\textbf{(a)} Success fraction $\phi_+$ as a function of wind-estimation memory time $\tau_w$ for Q-RL (green squares) and the spiral-search baseline (blue dashed line) in isotropic turbulence ($U = 0$). Vertical dashed lines mark the turbulent velocity correlation time $\tau_\mathrm{corr}$ (orange) and the plume intermittency correlation time $\tau_\mathrm{plume}$ (grey), both estimated independently from the DNS data (see main text). Inset: empirical detection-probability map $P_{\rm hit}$ for the isotropic environment, showing a radially symmetric plume structure centered on the source (star), with no preferred downstream direction.
\textbf{(b)} Box plots of the cumulative reward $R$ for Q-RL (green) at each tested memory value and for the spiral baseline (blue). The reward distributions mirror the pattern in panel~(a): intermediate memories yield higher and less variable rewards, while performance at the extremes degrades substantially, falling even below the spiral baseline.
\textbf{(c)} Box plots of the reciprocal of the normalized arrival time $T_{\rm min}/T$ conditioned on successful trajectories. 
\textbf{(d)} Representative trajectories in the isotropic environment starting from the green square), with the source marked by a star. An agent with the optimal memory $\tau_w = 5$ (yellow-red gradient) maintains frequent contact with the odor plume and reaches the source efficiently, tracking the local flow direction on the right timescale. An agent with a longer memory $\tau_w = 30$ (yellow-green gradient) reorients its actions along a slowly varying, uninformative wind estimate, loses contact with the plume, and fails to reach the source within the time horizon --- a direct visual illustration of the performance collapse at large $\tau_w$ seen in panels~(a)--(c). The spiral-search baseline (light-to-dark blue) explores space systematically but without any directional bias.
Results shown in panels~(a)--(c) are obtained from $5\cdot10^4$ distinct initial conditions.}
\label{fig:4}
\end{figure*}

To characterize these behavioral differences quantitatively, we compute the conditional mean-squared displacements with respect to the position of the last detection $(x_d, y_d)$, separated into the longitudinal component $(x - x_d)^2$ along the estimated wind direction and the transverse component $(y - y_d)^2$, both as functions of the time since last detection $\tau_d$ (Fig.~\ref{fig:3}(b)). The state-visitation distribution $P(\tau_d)$, shown in Supplementary Fig.~S2(a), reveals that values of $\tau_d \gtrsim 100$ are rarely encountered by either policy; the displacement analysis is therefore restricted to this functionally active range, where the statistics are well-sampled and representative of typical search behavior.
The two memory extremes produce strikingly different displacement profiles.
Short-memory agents ($\tau_w = 1$) show no significant casting with the transverse displacement remaining small throughout. The longitudinal displacement is similarly small at short $\tau_d$, with no discernible up- or downwind bias in the first few steps after a detection. Only at larger $\tau_d$ does the longitudinal component grow appreciably, a sign that the agent has drifted away while trying to re-establish contact with the plume without strong directional guidance. 
Despite this apparent trapping in the crosswind direction in the wind-relative frame, agents explore transversally to the mean wind (Fig.~\ref{fig:3}(c)), as confirmed by the diffusive growth of $\langle(y-y_d)^2\rangle$ measured with respect to the allocentric reference frame (Supplementary Fig.~S2(b)): when $\tau_w = 1$, rapid angular decorrelation of the wind estimate renders consecutive crosswind increments independent, so that lateral coverage emerges as a fluctuation-driven random walk rather than directed casting.
Conversely, long-memory agents ($\tau_w = 100$) develop a clear behavioral signature as $\tau_d$ increases: the transverse displacement grows substantially, reflecting active casting across the estimated wind direction, while the longitudinal displacement simultaneously oscillates, reflecting periodic backtracking against the estimated wind direction. The combination of lateral casting and downwind backtracking produces the geometrically structured, upwind-drifting spiral-like exploration that distinguishes this policy from the short-memory one.
For reference, the cast-and-surge baseline displayed in Fig.~\ref{fig:3}(b) shows a markedly different profile: a pronounced and dominant transverse component reflecting its stereotyped lateral casting response, combined with a longitudinal component that grows steadily in the upwind direction, reflecting the consistent upwind surging that is hardwired into the heuristic.
The three strategies thus represent qualitatively distinct ways of exploiting the same wind-relative reference frame, differing in the balance between casting, backtracking, and directed upwind motion.

The representative trajectories in Fig.~\ref{fig:3}(c) make all these features visually explicit (see also Supplementary Movies~S1-S3). Short-memory agents ($\tau_w = 1$) produce irregular, diffusive paths with no discernible large-scale structure.
Long-memory agents ($\tau_w = 100$) instead execute broad lateral sweeps combined with progressive downwind excursions between detections, resulting in extended looping paths that gradually spiral towards the source --- a direct visual counterpart of the anisotropic displacement growth in Fig.~\ref{fig:3}(b).
Despite these marked geometrical differences, both policies achieve comparable success rates, as shown in Fig.~\ref{fig:2}: the wind memory time does not affect how well the task is solved, but it does affect how success is achieved.

Strikingly, policies trained in the harder mild-wind environment transfer without loss to a stronger wind regime, while the reverse transfer degrades substantially for long memory times, establishing a clear asymmetry: training in the more challenging regime yields strategies that are robust to increases in mean-flow strength (Supplementary Fig.~S3).

\section{Optimal memory in isotropic turbulence}

We now turn to the isotropic limit ($U = 0$), where no preferred large-scale direction exists, the detection-probability map $P_\mathrm{hit}$ is radially symmetric around the source (inset of Fig.~\ref{fig:4}(a)), and the locally estimated wind direction $\bar{\bm{U}}$ carries no persistent directional bias. This regime provides a stringent test of the wind-relative framework: without a mean flow to anchor the agent's reference frame, wind-direction estimation becomes a noisier and more volatile process, and the value of temporal wind integration is no longer self-evident.

In this regime, because the cast-and-surge strategy performs poorly in the absence of a mean wind, we instead adopt spiral search as the benchmark policy~\cite{masson2009,barbieri2011}, whereby the agent follows an expanding spiral trajectory that is reset after each new detection (see Fig.~\ref{fig:4}(d) and Appendix C for details).
Q-RL policies outperform this baseline across a range of $\tau_w$ values, confirming that the wind-relative framework remains effective even in the complete absence of mean flow.
However, in sharp contrast to the mild-mean-wind regime, aggregate performance now depends strongly and non-monotonically on $\tau_w$ (Fig.~\ref{fig:4}(a,b)). Both the success fraction $\phi^+$ and the cumulative reward $R$ exhibit a clear optimum at intermediate wind memory times ($\tau_w \approx 3$--$7$): Q-RL outperforms the spiral baseline in this range, while performance at both extremes degrades substantially, falling even below the spiral baseline for the longest memory times tested. 
This picture is reinforced by the arrival-time distributions conditioned on successful trajectories (Fig.~\ref{fig:4}(c)). Agents with intermediate $\tau_w$ not only reach the source more reliably but also do so more efficiently, with arrival-time distributions shifted towards shorter values compared to both the spiral baseline and the performance at the extremes. The optimal $\tau_w$ therefore simultaneously maximizes reliability and efficiency, establishing the wind memory time as an active control parameter in this regime, rather than a passive tuning knob as in the mean-wind case.

The origin of this optimum lies in the temporal structure of the turbulent flow.
The velocity field in isotropic turbulence is correlated only over a finite time $\tau_\mathrm{corr}$, which we estimate from the spatially averaged autocorrelation function of the velocity signal. 
On the one hand, when $\tau_w \gg \tau_\mathrm{corr}$, the exponential kernel integrates over many statistically independent velocity samples, so that the resulting wind estimate reflects the accumulated history of uncorrelated fluctuations.
As a result, the agent's wind-relative reference frame effectively locks onto an arbitrary, uninformative heading, causing performance to degrade sharply and to fall even below the spiral baseline (see, e.g., Fig.~\ref{fig:4}(a)), which at least explores space systematically without any directional commitment.
On the other hand, when $\tau_w \ll \tau_\mathrm{corr}$, the policy tracks instantaneous turbulent fluctuations without fully exploiting the coherent directional structure present on timescales shorter than $\tau_\mathrm{corr}$: although instantaneous wind measurements still carry useful directional information, the agent reorients its actions too rapidly to accumulate a useful wind estimate, and performance is suboptimal.

An intermediate $\tau_w$ allows the agent to integrate velocity measurements over a window that is long enough to filter out incoherent noise yet short enough to track the locally coherent flow direction as it evolves. The qualitative differences between these regimes are illustrated by the representative trajectories in Fig.~\ref{fig:4}(d) (see also Supplementary Movies~S4-S6).
Supporting this interpretation, we estimate the correlation timescale of the odor plume by computing the autocorrelation function of the hits signal $h(t)$ --- i.e., the thresholded concentration signal recording the binary sequence of detection events --- at each spatial point and averaging its decay time over space; this plume-correlation timescale $\tau_\mathrm{plume}$ is found to be close to the optimal wind memory time ($\tau_\mathrm{plume} \approx 7$, with the performance optimum at $\tau_w \approx 5$).

Taken together, these results identify temporal wind integration as a regime-dependent resource whose value is set by the interplay between the intrinsic timescales of the turbulent velocity field and the spatial statistics of the plume.


\section{Conclusions}

We have shown that a minimal-state wind-relative reinforcement-learning agent can reliably locate odor sources in turbulent flows and consistently outperform physically motivated heuristic baselines across two complementary flow regimes. The agent requires no global reference frame, no prior knowledge of the environment, and no complex internal state: its only inputs are the time elapsed since the last odor detection and a locally filtered estimate of wind direction parameterized by the wind memory time $\tau_w$. That such a parsimonious representation suffices to produce robust performance across qualitatively different flow conditions highlights the effectiveness of learning-based methods in leveraging complex environmental structure that is otherwise difficult to specify through manual engineering.

A central finding is the regime-dependent role of the wind-estimation memory time $\tau_w$. 
In the presence of a mild mean wind, $\tau_w$ shapes the geometry of the search without affecting aggregate performance: different $\tau_w$ produce qualitatively distinct search patterns --- from diffusive, unstructured exploration at short $\tau_w$ to structured, spiral-like casting and backtracking at long $\tau_w$ --- yet all achieve comparable success rates, search times, and cumulative rewards.
This suggests that biological or robotic navigators operating in environments characterized by mean flow have substantial flexibility in how they integrate temporally varying wind cues, and that the characteristic duration of wind memory may be limited not primarily by navigational performance itself, but rather by other factors, such as the intrinsic timescales of the available mechanosensory processing pathways.

In isotropic turbulence, by contrast, $\tau_w$ becomes an active determinant of performance with a well-defined optimum, whose value is linked to the coherence timescales of both the velocity field and the odor plume. In fact, in this limit, the fluctuations themselves are the only available directional resource, and exploiting them requires matching the integration window to the environment's coherence time.
The timescale-matching principle identified in isotropic turbulence thus offers a complementary and testable prediction: the optimal $\tau_w$ is set by the environment rather than by agent-level parameters. Whether biological wind-memory timescales reflect a similar environmental matching or are instead shaped primarily by physiological constraints on mechanosensory processing remains an open intriguing question.

Our results also speak directly to the role of wind sensing in computational models of olfactory search. Several recent approaches employ richer internal states, recurrent architectures, or prior statistical knowledge of the environment~\cite{singh2023, heinonen2025, rando2025}. Our framework deliberately forgoes all of these, and yet achieves robust performance: the key ingredient is not a rich internal state but a locally estimated, temporally integrated wind direction. This has practical implications for robotic olfactory systems, where a single anemometer integrated over an appropriately chosen $\tau_w$ may provide most of the directional information needed for reliable source finding, without requiring complex state estimation or environmental mapping.

Several directions remain open for future work. Extending the framework to three-dimensional turbulence, richer sensory modalities, actuation delays and noise, energetic costs of locomotion, and continuous action spaces are natural steps towards greater biological and engineering realism. Most intriguingly, agents that can \emph{learn} $\tau_w$ online --- adapting their temporal integration window in response to environmental feedback --- would provide a direct test of the timescale-matching principle identified here, and move towards a fully adaptive olfactory search strategy.

\section*{Acknowledgements}
We thank A. Celani and A. Loisy for useful discussions.
This work was supported by the European Research Council (ERC) under the European Union’s Horizon 2020 research and innovation program (Grant Agreement Nos.\ 882340 and 101002724) and by the National Institute of Health under grant R01DC018789. We also acknowledge financial support under the National Recovery and Resilience Plan (NRRP), Mission 4, Component 2, Investment 1.1, Call for tender No. 104 published on 2.2.2022 by the Italian Ministry of University and Research (MUR), funded by the European Union — NextGenerationEU —
Project Title Equations informed and data-driven approaches for collective optimal search in complex flows (CO-SEARCH), Contract 202249Z89M. --- CUP B53-D23003920006 and E53-D23001610006, and by a France 2030 support managed by the Agence Nationale de la Recherche, under the reference ANR-23-PEIA-0004 (PDE-AI project).
This work represents only the views of the authors; the European Research Council Executive Agency and the other funding agencies are not responsible for any use that may be made of the information it contains.

\section*{Data availability}
The code used to reproduce the results of this study is publicly available on \href{https://github.com/LorenzoPiro95/Qnav-windEstimate}{GitHub}. Data can be made available upon reasonable request to the corresponding author.


\appendix

\section{DNS of turbulent flow and scalar transport}

Turbulent environments are generated by numerically integrating the
two-dimensional incompressible Navier--Stokes equations coupled to an
advection--diffusion equation for a passive scalar field $\theta$ emitted
by a point-like source:
\begin{subequations}
\begin{align}
	&\partial_t\bm{u} +\bm{u\cdot \nabla u} = -\bm{\nabla} P +\nu\nabla^2 \bm{u} - \alpha \bm{u} + \bm{f}, \;       \label{eq_NS}
\\
	&\partial_t\theta + \bm{u}\bm{\cdot \nabla}\theta = \kappa\nabla^2 \theta -\frac{\theta}{\mathcal{T}} + R\tilde{\delta}(\bm{x}-\bm{x}_{s}).
    \label{eq_scal}
\end{align}
\end{subequations}
Both equations are integrated using a pseudo-spectral code with $2/3$ dealiasing in a square $2\pi\times 2\pi$ domain with periodic boundary conditions on a $N^2=1024^2$ grid.

\begin{figure}[t!]
\centering
\includegraphics[width=0.85\columnwidth]{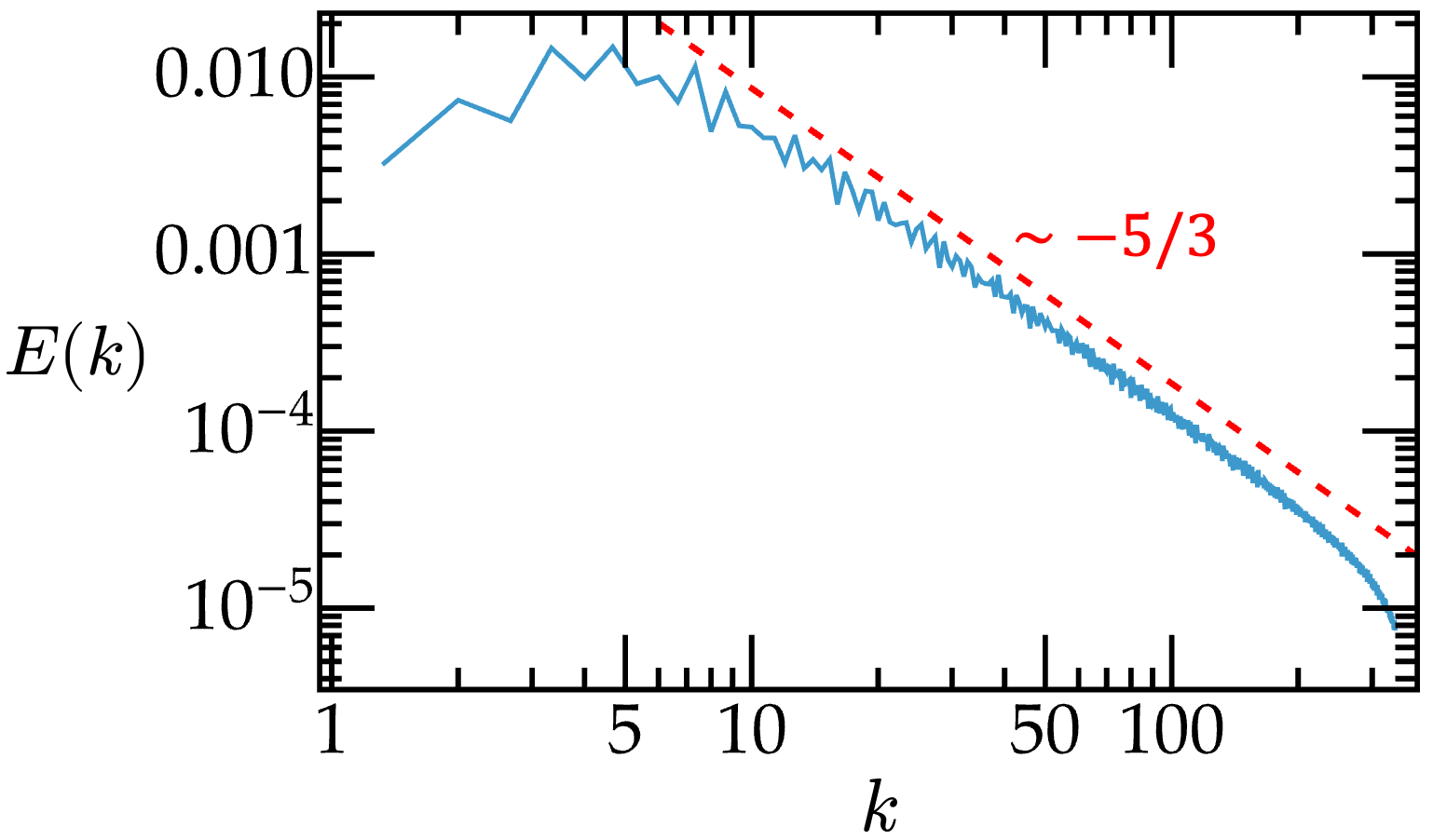}
\caption{\justifying \textbf{Kinetic energy spectrum of the turbulent velocity field from DNS.} Time-averaged kinetic energy spectrum $E(k)$ as a function of wavenumber $k$, computed from the two-dimensional DNS in the statistically stationary state. The spectrum exhibits a clear power-law scaling range consistent with $E(k) \propto k^{-5/3}$ (dashed red line) over more than a decade in wavenumber, confirming that the simulated flow operates in the inverse energy cascade regime of two-dimensional turbulence. The deviation from the power law at large $k$ reflects the onset of hyperviscous dissipation, which removes energy at the small scales. The spectrum is the same across all three flow regimes ($U/u_\mathrm{rms} = 0,\,1,\,2$), consistent with the fact that the mean wind is imposed via the $k=0$ mode and does not alter the statistical properties of the fluctuating velocity field.}
\label{fig:spectrum}
\end{figure}

In Eq.~\eqref{eq_NS}, $P$ is the pressure enforcing incompressibility ($\bm{\nabla\cdot u}=0$), $\nu$ is the kinematic viscosity, and $\bm f$ is a Gaussian, zero-mean, time-uncorrelated forcing acting at small scales $\ell_f\approx 0.03$, which drives an inverse energy cascade, with a Kolmogorov spectrum $E(k)\propto k^{-5/3}$ towards the large scales (see Fig.~\ref{fig:spectrum}).
This implies that the velocity field is spatially rough but lacks the intermittency characteristic of three-dimensional turbulence~\cite{boffetta2012}. Energy accumulated at large scales is removed by the linear friction term $-\alpha\bm{u}$, whose coefficient $\alpha$ sets the large-scale dissipation.
A mean wind of intensity $\bm U=(U,0)$ is imposed by fixing the $k=0$ mode of the velocity field; by Galilean invariance, this does not alter the statistical properties of the velocity field. The viscous term is replaced by an order-8 hyperviscous operator, as customary in DNS of two-dimensional turbulence~\cite{boffetta2012}.

The simulations are run until both the velocity and the scalar fields reach a statistically stationary state, characterized by a root-mean-square velocity $u_{\rm rms}\approx 0.4$, independent of the mean flow, and an integral scale $L\approx 1$. The small-scale characteristic time is estimated as $\tau_s=\Omega^{-1/2}\approx 0.05$, where $\Omega=\langle \omega^2\rangle/2$ is the mean enstrophy of the flow. 

Three flow regimes are considered: a mild--mean-wind case ($U/u_{\rm rms}=1$), a strong-mean-wind case ($U/u_{\rm rms} = 2$), and an isotropic case ($U = 0$). In each regime, snapshots of the velocity and scalar fields are stored at time intervals $\delta t$ such that $\max\{U, u_{\rm rms}\}\,\delta t \approx 2\,\delta x$, where $\delta x=2\pi/N$ is the mesh resolution, ensuring that the fastest flow structures are advected by at most two mesh cells between consecutive stored frames. A total of $2\cdot10^4$ snapshots are stored per regime.

Equation~\eqref{eq_scal} governs the passive scalar field, which is damped at small scales by diffusion with diffusivity $\kappa$ and decays over a timescale $\mathcal{T}$. The scalar is emitted by a point-like source located
at $\bm x_{s}=(512,512)\delta x$ for the isotropic case, $\bm x_{s}=(300,512)\delta x$ for the mild--mean-wind case, and in $\bm x_{s}=(150,512)\delta x$ for strong mean wind. The point-like source is numerically represented as a Gaussian distribution with standard deviation $3\delta x$ and source emission rate $R$.

To prevent the scalar field from re-entering the domain through the periodic boundaries, a penalization mask~\cite{schneider2005numerical}, smoothed by a hyperbolic tangent, is applied near the domain edges, on a strip of width $40\delta x$,
which gradually removes the scalar in the boundary region.

\begin{table}[h!]
\centering
\caption{Numerical parameters of the DNS and coarse-grained fields. Parameters are common to all three flow regimes ($U/u_\mathrm{rms} = 0,\,1,\,2$) unless otherwise noted. All quantities are given in simulation units.}
\label{tab:dns_params}
\begin{ruledtabular}
\begin{tabular}{lcc}
\textbf{Parameter} & \textbf{Symbol} & \textbf{Value} \\
\hline
\multicolumn{3}{l}{\textit{DNS fields ($1024^2$ grid)}} \\
\hline
Grid resolution          & $N^2$                      & $1024^2$ \\
Domain size              & $\mathcal{L}$              & $2\pi \times 2\pi$ \\
Mesh size                & $\delta x = 2\pi/N$        & $6.13\times10^{-3}$ \\
RMS velocity             & $u_\mathrm{rms}$           & $0.4$ \\
Integral scale           & $L$                        & $1$ \\
Small-scale time         & $\tau_s = \Omega^{-1/2}$   & $0.05$ \\
Forcing scale            & $\ell_f$                   & $0.03$ \\
Emission rate            & $R$                        & $0.15$ \\
Source width             & $\sigma_s$                 & $3\,\delta x$ \\
Simulation time step     & $dt$                       & $2\times10^{-4}$ \\
Snapshot interval        & $\delta t$                 & $0.032$ ($0.016$)$^\dagger$ \\
Total snapshots stored   &                            & $2\cdot10^4$ \\
\hline
\multicolumn{3}{l}{\textit{Coarse-grained fields ($128^2$ grid)}} \\
\hline
Coarse grid resolution   & $M^2$                      & $128^2$ \\
Effective mesh size      & $\Delta x = 8\,\delta x$   & $0.049$ \\
Agent decision interval  & $\Delta t = 4\,\delta t$   & $0.128$ ($0.064$)$^\dagger$ \\
Agent speed              & $v_x = \Delta x/\Delta t$  & $0.38$ ($0.77$)$^\dagger$ \\
Source detection radius  & $r_s=2\Delta x$            & $0.098$ \\     
\end{tabular}
\end{ruledtabular}
\begin{flushleft}
{\small $^\dagger$ Values in parentheses refer to the strong mean wind
regime ($U/u_\mathrm{rms}=2$), where the faster advective timescale
requires a shorter snapshot interval.}
\end{flushleft}
\end{table}

The numerical parameters for all settings are summarized in Table~\ref{tab:dns_params}, and Supplementary Movies~S7 and~S8. illustrate the time evolution of the scalar field with and without mean wind.

The stored high-resolution fields are coarse-grained from the original $N^2 = 1024^2$ grid onto a reduced $M^2 = 128^2$ grid by averaging both the velocity and scalar concentration over non-overlapping $8\times 8$ blocks of fine-grid cells, yielding a coarsening factor of $8$ in each spatial direction.
The coarse-grained fields are further subsampled in time by retaining one frame every four snapshots, giving an effective agent decision interval of $\Delta t = 4\,\delta t$. From these coarse-grained fields, we compute the empirical detection-probability map $P_{\rm hit}(\bm{x})$ for each flow regime, defined as the time-averaged probability that the local scalar concentration exceeds the threshold $c_{\rm thr}$.

\section{Agent dynamics and reinforcement learning protocol}

\paragraph*{Agent dynamics.}
At each decision time step $\Delta t$, the agent samples the local scalar concentration $c(\bm{x}, t)$ and velocity field $\bm{u}(\bm{x}, t)$ at its current position by linear interpolation of the coarse-grained DNS fields (see above), and moves by $\Delta x$ in the chosen wind-relative direction, defining a constant speed $v = \Delta x/\Delta t$. Here and throughout, we work in units of the agent decision interval and step size, setting $\Delta t = \Delta x = 1$ without loss of generality; the corresponding values in simulation units are reported in Table~\ref{tab:dns_params}.

A detection event, or hit, is registered at time $t$ if the local concentration exceeds a fixed threshold,
\begin{equation}
    h(t) = \mathbf{1}\bigl[c(\bm{x}(t), t) > c_\mathrm{thr}\bigr],
\end{equation}
where $c_\mathrm{thr}=0.3\approx 6\bar{c}$ is the prescribed agent's sensitivity (see Fig.~\ref{fig:1}(a)).
At each decision step, the agent also updates its internal wind estimate $\bar{\bm{U}}$ by exponentially filtering the instantaneous local wind direction $\bm{u}$,
\begin{equation}
    \label{eq_expKernel}
    \bar{\bm{U}}(t+\Delta t) =
    (1 - \alpha_M)\bar{\bm{U}}(t)
    + \alpha_M\,\bm{u}(t), \quad \alpha_M=1-e^{-\Delta t/\tau_w} \, ,
\end{equation}
where $\tau_w$ is the wind memory time.
The normalized estimate $\hat{\bm{U}} = \bar{\bm{U}}/\|\bar{\bm{U}}\|$ defines the agent's wind-relative reference frame at each step.

\paragraph*{State space and action space.}
The agent's internal state is the elapsed time since the last hit, discretized into $T_H=500$ bins,
\begin{equation}
    \tau_d^{(k)} \in \{0, 1, 2, \ldots, T_H-1\},
\end{equation}
where $\tau_d$ increases by one unit at each decision time step and is reset to zero upon each new detection. The state space thus spans the full episode horizon, with the largest bin $\tau_d = T_H-1$ absorbing all detection-free intervals that reach the time limit.
At each decision step, the agent selects one of four discrete actions $a \in \mathcal{A} = \{\uparrow, \downarrow, \leftarrow, \rightarrow\}$, corresponding to the four cardinal directions. The action set is then rotated at each step to align with the current wind estimate $\hat{\bm{U}}$, so all actions are defined in the wind-relative frame.

\paragraph*{Reward function.}
Policies are trained by maximizing the cumulative discounted reward
\begin{equation}
    R = \sum_{t=0}^{T} \gamma^t\, r_t,
\end{equation}
where $T$ is the time step at which the source is reached, $\gamma = 0.998$ is the discount factor, and the immediate reward is
\begin{equation}
    r_t =
    \begin{cases}
        +1 & \text{if the source is reached at step } t = T, \\
        \gamma - 1 & \text{otherwise.}
    \end{cases}
\end{equation}
The source is considered reached when the agent comes within a distance
$r_s = 2\,\Delta x$ of the source location $\bm{x}_s$.
This choice of running reward $r_t = \gamma - 1 < 0$ ensures that the cumulative reward is maximized simultaneously by (i) reaching the source reliably and (ii) reaching it as quickly as possible, since faster trajectories accumulate fewer negative steps before collecting the terminal bonus. Episodes that do not reach the source within the finite time horizon $T_H$ are terminated without a terminal reward; such trajectories are classified as lost. 

\paragraph*{Tabular Q-learning.}
The Q-function $Q(\tau_d, a)$, representing the maximum expected cumulative discounted reward starting from state $\tau_d$ and taking action $a$, is estimated by tabular Q-learning via the Bellman update~\cite{sutton2018}
\begin{equation}
    Q(\tau_d, a) \leftarrow Q(\tau_d, a) + \alpha_n \Bigl[r + \gamma \max_{a'} Q(\tau_d', a') - Q(\tau_d, a)\Bigr],
\end{equation}
where $\tau_d'$ is the state at the next step, $r$ is the immediate reward received, and $\alpha_n$ is the learning rate at the $n$-th episode.
The optimal policy is the greedy readout $\pi^*(\tau_d) = \arg\max_a\, Q(\tau_d, a)$.
Exploration during training follows an $\varepsilon$-greedy schedule~\cite{sutton2018}: at each step, the agent selects a uniformly random action with probability $\varepsilon_n$ and the current greedy action otherwise.
Both the learning rate $\alpha_n$ and the exploration parameter $\varepsilon_n$ decay exponentially over $N_\mathrm{train} = 10^4$ training episodes,
\begin{equation}
    \alpha_n = \alpha_0 \left(\frac{\alpha_f}{\alpha_0}
    \right)^{n / N_\mathrm{train}},
    \qquad
    \varepsilon_n = \varepsilon_0 \left(\frac{\varepsilon_f}
    {\varepsilon_0}\right)^{n / N_\mathrm{train}},
\end{equation}
with initial values $\alpha_0 = 0.1$, $\varepsilon_0 = 1.0$ and final values $\alpha_f = \varepsilon_f = 10^{-4}$.

\paragraph*{Training protocol.}
At each training episode, a DNS snapshot is selected uniformly at random among the first 2000 frames of a stored sequence of $2500$ independent flow realizations, with consecutive snapshots being separated by a time interval equal to the agent's decision time $\Delta t$; this ensures that the full episode horizon $T_H=500$ is contained within the available data.
Then, $N_\mathrm{batch} = 256$ agents are initialized in parallel at positions drawn uniformly from the set of locations where the instantaneous scalar concentration exceeds $c_\mathrm{thr}$ in that snapshot, ensuring that every agent registers a hit upon initialization and begins the episode in the well-defined state $\tau_d = 0$.
The Q-table is updated after each episode using the transitions accumulated by all 256 agents, providing a diverse and decorrelated batch of experience at each update as well as statistically robust Q-value estimates. Convergence is monitored by the plateau of the success fraction $\phi_+$, the normalised arrival time $T_\mathrm{min}/T$, and the cumulative reward $R$ (Fig.~\ref{fig:1}(d)). Performance at test time is evaluated on an independent set of $2500$ DNS snapshots, disjoint from those used during training, to ensure that reported metrics reflect generalization to unseen flow realizations rather than memorization of training conditions. Pseudo-code for the full training loop is provided in the Supplementary Material.\\

\section{Baseline strategies}

Two heuristic baselines are used for benchmarking; both are evaluated over the same ensembles of initial conditions and flow realizations used to assess the learned policies. 

\paragraph*{Cast-and-surge.}
Cast-and-surge~\cite{balkovsky2002} is a biologically inspired heuristic modeling the upwind-casting behavior observed in flying insects~\cite{murlis1992,carde2021}. 
Originally formulated for a known, fixed mean wind direction, we adapt it here to our setting by replacing the actual wind direction with the agent's internal wind estimate, updated online during the search (see Eq.~\eqref{eq_expKernel}). The original cast-and-surge formulation is asymptotically recovered in the limit of large values of $\tau_w$.
The agent moves at constant speed and selects its heading according to a hardwired policy that depends solely on its internal clock $\tau_d$, the time elapsed since the last odor detection.
Upon a detection event (hit, $\tau_d = 0$), the policy prescribes a \emph{surge}: the agent moves in the upwind direction.
Once the plume is lost, i.e., when no hit is registered, the policy switches to a \emph{casting} phase, in which the agent moves in a zig-zag fashion, always transversally to the current estimated mean wind direction, with turning times that increase linearly with the time from the last odor detection. The \emph{cast angle} $\theta_c \in [0, \pi/2]$, equal to half the aperture of the casting cone, therefore controls the width of the crosswind sweep: smaller values result in a more strongly upwind-directed trajectory with minimal crosswind deviations, whereas values approaching $\pi/2$ yield a more cautious and exploratory search pattern characterized by reduced net upwind displacement.
At every time step, the prescribed heading is rotated into the wind-relative reference frame defined by the agent's current wind estimate $\bar{\bm{U}}$, which is computed via the same exponential memory kernel with characteristic time $\tau_w$ used by the Q-RL agent.
The cast angle $\theta_c$ is the sole free parameter of this heuristic; it is tuned independently for each value of $\tau_w$ by grid search over the values $\theta_c \in \{ 30^{\circ}, 45^{\circ}, 60^{\circ}, 70^{\circ}, 80^{\circ} \}$ so as to maximize the mean cumulative reward $\langle R \rangle$ computed along the agent's trajectories in the mild-mean-wind environment (see inset of Fig.~\ref{fig:2}(a)), ensuring that the baseline is as competitive as possible at every operating condition.

\paragraph*{Spiral search.}
Spiral search~\cite{masson2009,barbieri2011} is used as the baseline in the isotropic regime ($U=0$), where no mean-wind direction is available to anchor an upwind-casting strategy. 
The agent moves at constant speed and follows a hardwired policy that depends solely on $\tau_d$: starting from the position of the most
recent detection event, it executes an outward Archimedean spiral by
sequentially traversing arms of increasing length along the four cardinal directions.
After each completed quarter-turn, the arm length is incremented by a fixed amount $\delta \ell$, so that the covered area expands monotonically with time. Whenever a new detection is registered ($\tau_d = 0$), the spiral is reset and restarts centered on the new hit position (see the trajectory at the bottom of Fig.~\ref{fig:4}(d) for an illustrative example).
The sole free parameter is the \emph{spiral growth rate} $\delta\ell$, which controls how much the arm length is incremented after each quarter-turn.
It is tuned by grid search over the values $\delta\ell \in \{1, 2, 3, 4, 5, 6, 7, 8, 9\}$ to maximize $\langle R \rangle$ in the isotropic environment, using the same optimization criterion adopted for the cast-and-surge baseline.


\bibliographystyle{apsrev4-2}

%

\clearpage

\onecolumngrid

\section*{Supplemental Material}

\subsection{Supplementary figures}

\begin{figure*}[ht!]
\centering
\includegraphics[width=\textwidth]{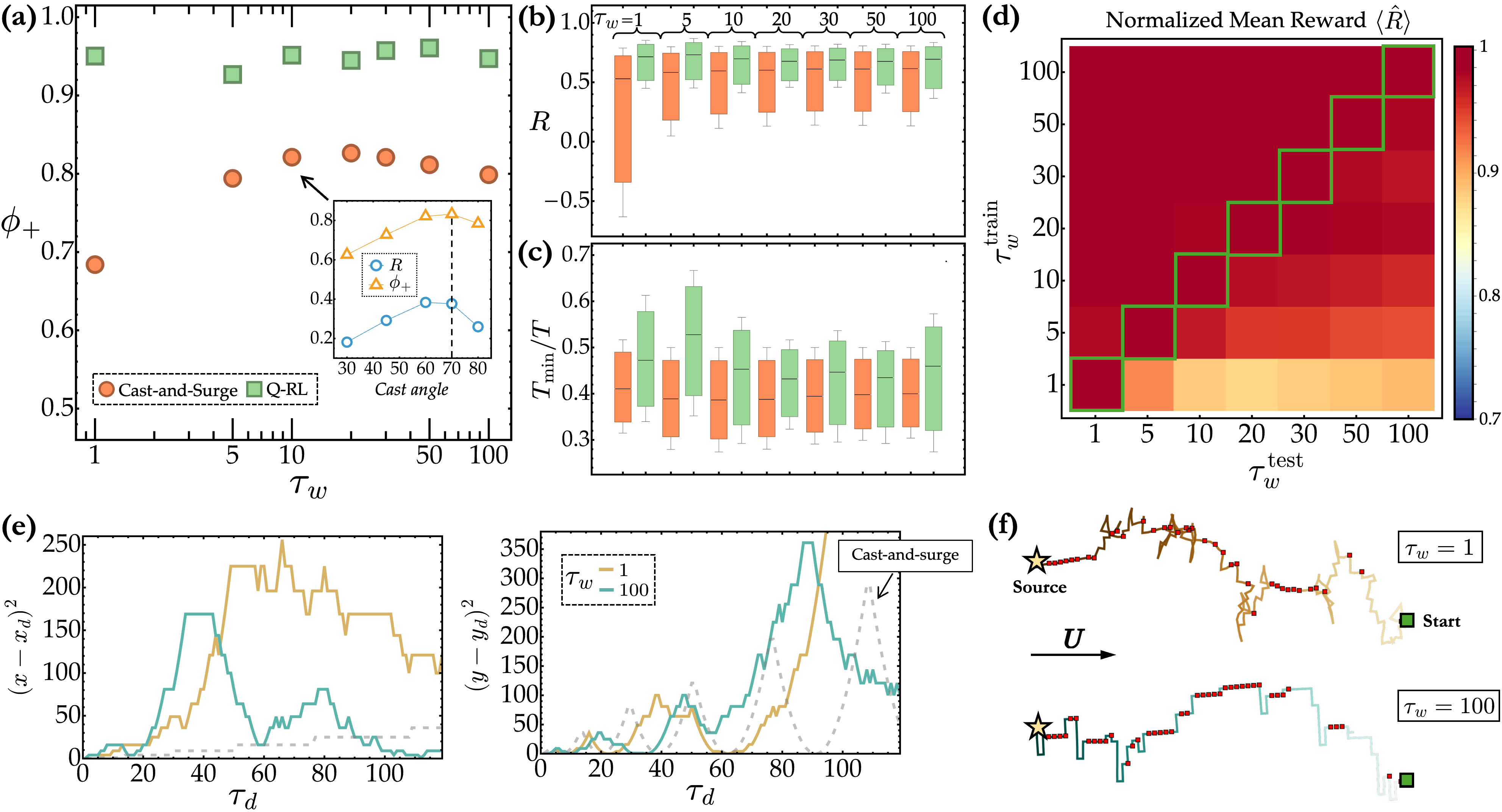}
\caption{\justifying \textbf{Performance of learned policies versus cast-and-surge in the stronger mean wind regime ($U/u_\mathrm{rms} = 2$).}
\textbf{(a)} Success fraction $\phi_+$ as a function of wind-estimation memory time $\tau_w$ for Q-RL (green squares) and cast-and-surge (red circles). Q-RL consistently outperforms the baseline across all $\tau_w$, with $\phi_+ \gtrsim 0.93$, while cast-and-surge achieves $\phi_+ \approx 0.8$ at its optimally tuned casting angle (inset).
The near-flat dependence on $\tau_w$ confirms that a strong mean flow renders the learned policy largely insensitive to wind-estimation memory.
\textbf{(b)} Box plots of the cumulative reward $R$ for Q-RL (green) and cast-and-surge (red). Q-RL systematically achieves higher, less variable rewards across all $\tau_w$.
\textbf{(c)} Box plots of $T_\mathrm{min}/T$ conditioned on successful trajectories. Arrival-time distributions of Q-RL and cast-and-surge are broadly comparable, indicating that Q-RL's advantage lies primarily in its higher success rate rather than faster navigation.
\textbf{(d)} Cross-testing matrix of the normalized mean reward $\langle\hat{R}\rangle$ for policies trained at $\tau_w^\mathrm{train}$ (rows) and deployed at $\tau_w^\mathrm{test}$ (columns), without retraining. Rewards are normalized by the matched diagonal values (green outline), i.e., the performance of each policy when tested with the same memory used during training. 
Except for $\tau_w^\mathrm{train} = 1$, the matrix is close to unity off-diagonal, demonstrating broad transfer robustness; the strong mean flow makes $\tau_w$ a secondary factor for all but the shortest memory.
\textbf{(e)} Conditional mean-squared displacements with respect to the last detection position $(x_d, y_d)$: longitudinal $(x-x_d)^2$ (left) and transverse $(y-y_d)^2$ (right) as functions of $\tau_d$, for $\tau_w = 1$ (gold), $\tau_w = 100$ (teal), and cast-and-surge (dashed grey). Both memory extremes show similar profiles, with dominant transverse casting and infrequent longitudinal reversals, in contrast to the mild-wind regime where the two extremes differ markedly.
\textbf{(f)} Representative trajectories for $\tau_w = 1$ (top, gold) and $\tau_w = 100$ (bottom, teal) from the same initial position (green square); source marked by a star, mean wind by $\boldsymbol{U}$, odor hits by red squares. Both agents perform casting with infrequent wind-direction reversals, consistent with panel (e), confirming that, in the strong mean wind regime, the learned policy does not strongly adapt to wind-estimation memory.}
\label{fig:wind2}
\end{figure*}

\begin{figure*}[ht!]
\centering
\includegraphics[width=\textwidth]{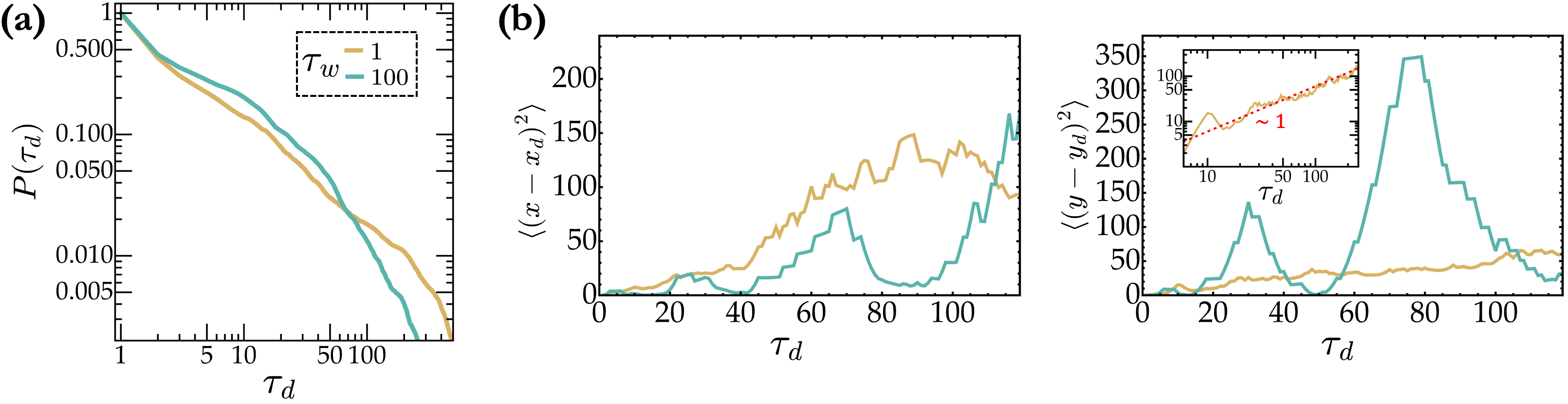}
\caption{\justifying \textbf{Mean-squared-displacement analysis of learned strategies in the mild mean wind regime ($U/u_\mathrm{rms} = 1$).}
Results are shown for $\tau_w = 1$ (gold) and $\tau_w = 100$ (teal).
\textbf{(a)} State-visitation probability $P(\tau_d)$ normalized at its value at $\tau_d=0$ as a function of the time $\tau_d$ elapsed since the last odor detection, on a log-log scale. For both memory times, $P(\tau_d)$ decays rapidly and falls below $\sim 1$--$2\%$ for $\tau_d \gtrsim 100$, confirming that long detection-free intervals are rarely visited; the displacement analysis in panels (b)--(c) is therefore restricted to this functionally active range.
\textbf{(b)} Conditional mean-squared displacements in the allocentric reference frame averaged over different flow realizations and initial conditions. Top: longitudinal component $\langle(x-x_d)^2\rangle$. Bottom: transverse component $\langle(y-y_d)^2\rangle$. For $\tau_w = 100$, the motion is strongly anisotropic at large $\tau_d$, with enhanced transverse excursions reflecting active casting and longitudinal oscillations reflecting periodic backtracking. For $\tau_w = 1$, the transverse component grows approximately linearly with $\tau_d$ (inset, log-log scale; dashed red line indicates diffusive scaling $\sim \tau_d^1$). This diffusive behavior admits a transparent analytical interpretation. When $\tau_w = 1$, the wind estimate $\bar{\bm{U}}$ tracks instantaneous turbulent fluctuations and decorrelates rapidly between successive steps. Each action is therefore rotated by a nearly independent random angle, so that the crosswind component of consecutive displacements is uncorrelated: the transverse displacement accumulates as a sum of independent random increments, yielding diffusive growth $\langle(y - y_d)^2\rangle \sim \tau_d$ by the central limit theorem. The short-memory agent has thus learned to exploit the fast angular decorrelation of the flow to execute a random walk in the crosswind direction, achieving lateral coverage through fluctuation-driven diffusion rather than directed casting.}
\label{fig:MSD_analysis_W1}
\end{figure*}

\begin{figure*}[ht!]
\centering
\includegraphics[width=0.8\textwidth]{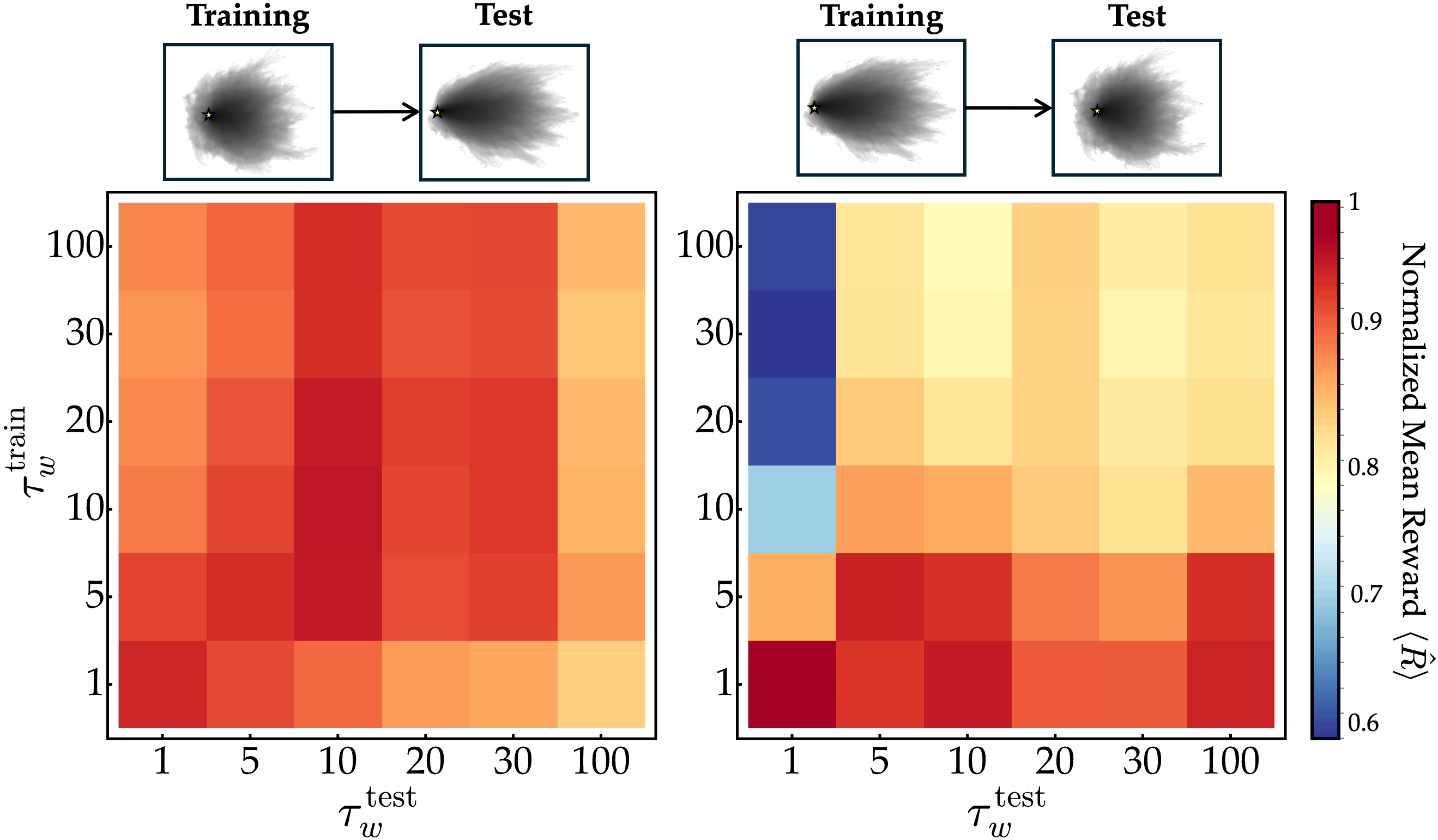}
\caption{\justifying \textbf{Cross-environment transfer of learned policies between the mild ($U/u_\mathrm{rms} = 1$) and strong ($U/u_\mathrm{rms} = 2$) mean wind regimes.}
Each matrix shows the normalized mean reward $\langle \hat{R} \rangle$ when a policy trained in one flow environment is deployed in a different one, as a function of the wind-estimation memory time used during training ($\tau_w^\mathrm{train}$, rows) and during testing ($\tau_w^\mathrm{test}$, columns).
Each entry is normalized by the performance achieved when the same policy is tested in the environment it was trained on, so that values below unity directly quantify the performance loss due to environment mismatch.
The schematics above each matrix indicate the direction of transfer: the training plume (left thumbnail) is transported to the test plume (right thumbnail).
\textbf{(Left)} Transfer from the mild-wind environment ($U/u_\mathrm{rms} = 1$, training) to the strong-wind environment ($U/u_\mathrm{rms} = 2$, test). The matrix is uniformly warm-coloured and close to unity across all $(\tau_w^\mathrm{train}, \tau_w^\mathrm{test})$ pairs, demonstrating that policies learned in the harder, mild-wind regime transfer essentially without loss to the easier, strong-wind regime.
This asymmetry is physically natural: a policy trained under a weaker mean flow, where wind-direction estimation is a genuine challenge and turbulent fluctuations are comparably strong, develops a robust search strategy that remains effective when a more reliable directional cue becomes available.
\textbf{(Right)} Transfer from the strong-wind environment ($U/u_\mathrm{rms} = 2$, training) to the mild-wind environment ($U/u_\mathrm{rms} = 1$, test). Performance degrades substantially for policies trained at large $\tau_w^\mathrm{train} \geq 10$. Policies trained at short memory ($\tau_w^\mathrm{train} \leq 5$) transfer more successfully, recovering near-diagonal performance at small $\tau_w^\mathrm{test}$.
This degradation reflects the fact that the strong-wind environment is an easier task: strategies learned there over-exploit the reliable mean-flow direction and rely on long-memory wind estimates to navigate, developing a policy that is too rigid to cope with the stronger directional uncertainty
and larger turbulent fluctuations encountered in the mild-wind regime.
Together, the two panels establish a clear asymmetry in cross-environment generalization: for practical deployment of olfactory navigation agents across variable flow conditions, training in the more demanding, lower wind-to-turbulence-ratio regime is the preferable strategy.}
\label{fig:crosstest_winds}
\end{figure*}

\section{Details on the Q-learning algorithm} 

\begin{algorithm}[H]
\caption{Q-learning for olfactory search: the input comprises concentration $c(\bm{x}, t)$ and velocity fields $\bm{v}(\bm{x}, t)$ from DNS data, source position $\bm{x}_s$ and radius $r_s$, episode time horizon $T_H$, discount factor $\gamma$, wind-estimation memory time $\tau_w$, number of training episodes $N_{\rm train}$, initial and final learning and exploration rates $\alpha_0, \alpha_f$, $\varepsilon_0, \varepsilon_f$.}
\begin{algorithmic}[1]

\State \textbf{Initialize}
\State $Q(s, a) \gets -1$ \quad $\forall\, s \in \{0,\dots,T_H-1\},\; a \in \mathcal{A}=\{\bm{e}_x, -\bm{e}_x, \bm{e}_y, -\bm{e}_y\}$ \Comment{Initialize Q-table}
\State \State $B = 256$, \, $\alpha_k = \alpha_0\!(\alpha_f/\alpha_0)^{\!k / N_{\rm train}}$, \, $\varepsilon_k = \varepsilon_0 \!(\varepsilon_f/\varepsilon_0)^{\!k / N_{\rm train}}$ \Comment{Agents running in parallel per episode and decaying schedules}

\Statex

\For{each episode $k = 0, 1, \dots, N_{\rm train}$}
    \State Draw random time index $t_0$ uniformly from valid range
    \For{each agent $i = 1, \dots, B$ \textbf{in parallel}}
        \State $\bm{x}^{(i)}(0) \gets$ \Call{SampleInitialCondition}{$c,\, c_{\rm thr},t_0$}
        \State $s^{(i)} \gets$ \Call{UpdateState}{$c(\bm{x}^{(i)}(0),\, t_0)$} \Comment{Encode initial observation into discrete state}
        \State $\bm{\mu}^{(i)} \gets$ \Call{InitFlowMemory}{$\bm{v},\, \bm{x}^{(i)}(0),\, t_0,\, \tau_w$} \Comment{Initialize exponential moving average (EMA) of flow velocity}
    \EndFor

    \For{each time step $t = 0, 1, \dots, T_H - 1$}

        \For{each active agent $i = 1, \dots, B$ \textbf{in parallel}}

            \State $a^{(i)} \gets$ \Call{SelectAction}{$Q,\, s^{(i)},\, \varepsilon_k$} \Comment{$\varepsilon$-greedy action selection}
            \State $\bm{\mu}^{(i)} \gets (1 - \alpha_M)\,\bm{\mu}^{(i)} + \alpha_M\,\bm{v}(\bm{x}^{(i)}(t), t_0^{(i)}+t)$, \quad $\alpha_M = 1 - e^{-1/\tau_w}$ \Comment{Update flow EMA}
            \State $\bm{x}^{(i)}(t+1) \gets \bm{x}^{(i)}(t) + R\!\left(\mathrm{atan2}(\mu^{(i)}_y, \mu^{(i)}_x)\right){a^{(i)}}$ \Comment{Rotate action along mean flow and move}

            \If{$|\bm{x}^{(i)}(t+1) - \bm{x}_s|_1 \leq r_s$}
                \State $r^{(i)} \gets 1$, \quad $s^{(i)\prime} \gets -1$ \Comment{Terminal reward and state upon source found}
            \Else
                \State $r^{(i)} \gets \gamma - 1$ \Comment{Immediate reward}
                \State $s^{(i)\prime} \gets$ \Call{UpdateState}{$c\!\left(\bm{x}^{(i)}(t+1),\, t_0^{(i)} + t + 1\right)$} \Comment{Sample concentration and update state}
            \EndIf

        \EndFor

        \State \Call{UpdateQ}{$Q,\, \{s^{(i)}\},\, \{a^{(i)}\},\, \{s^{(i)\prime}\},\, \{r^{(i)}\},\, \alpha_k,\, \gamma$} \Comment{Single Q-table updated with all $B$ transitions}
        \State $s^{(i)} \gets s^{(i)\prime}$ \quad $\forall\, i$ with $s^{(i)\prime} \neq -1$

        \If{$s^{(i)\prime} = -1$ $\forall i$} \textbf{break} \EndIf \Comment{All agents have reached the source}

    \EndFor

\EndFor

\Statex

\Function{SampleInitialCondition}{$c,\, c_{\rm thr},t_0$}
    \State $\mathcal{X}_0 \gets \{\bm{x} : c(\bm{x}, t_0) > c_{\rm thr}\} \setminus \mathcal{B}(\bm{x}_s, r_s)$ \Comment{Valid positions above threshold, away from source}
    \State Draw $\bm{x}(0)$ uniformly from $\mathcal{X}_0$
    \State \Return $\bm{x}(0),\, t_0$
\EndFunction

\Statex

\Function{InitFlowMemory}{$\bm{v},\, \bm{x},\, t_0,\, 
\tau_w$}
    \State $\bm{\mu} \gets \bm{v}(\bm{x},\, t_0 - \tau_w)$
    \For{$\tau = t_0 - \tau_w + 1, \dots, t_0$}
        \State $\bm{\mu} \gets (1 - \alpha_M)\,\bm{\mu} + \alpha_M\,\bm{v}(\bm{x},\, \tau)$ \Comment{Warm-up EMA over past $M$ samples}
    \EndFor
    \State \Return $\bm{\mu}$
\EndFunction

\Statex

\Function{UpdateState}{$h$}
    \State \Return $s \gets \begin{cases} 0 & h \geq c_{\rm thr} \\ s + 1 & \text{otherwise} \end{cases}$ \Comment{Odor detection / no odor}
\EndFunction

\Statex

\Function{SelectAction}{$Q,\, s,\, \varepsilon$}
    \State \Return $a \gets \begin{cases} \sim \mathrm{Uniform}(\mathcal{A}) & u < \varepsilon, \quad u \sim \mathrm{Uniform}(0,1) \\ \arg\max_{a'} Q(s, a') & \text{otherwise} \end{cases}$ \Comment{Explore: random action / exploit: greedy action}
\EndFunction

\Statex

\Function{UpdateQ}{$Q,\, \{s^{(i)}\},\, \{a^{(i)}\},\, \{s^{(i)\prime}\},\, \{r^{(i)}\},\, \alpha,\, \gamma$}
    \State $Q(s^{(i)}, a^{(i)}) \gets (1-\alpha)\,Q(s^{(i)}, a^{(i)}) + \alpha\bigl[r^{(i)} + \gamma\,  \Theta(s^{(i)\prime})\max_{a'} Q(s^{(i)\prime}, a')\bigr]$ \quad $\forall\, i : s^{(i)} \geq 0$ \Comment{Tabular Q-learning update}
\EndFunction

\end{algorithmic}
\end{algorithm}

\section{Captions of Supplementary Movies}

\textbf{Supplementary Movie~S1 $|$ Representative trajectory of the Q-RL agent in the mild mean wind regime ($U/u_\mathrm{rms} = 1$, $\tau_w = 1$).}
The agent (gold trajectory) starts at the red square and navigates toward the odor source (green circle), whose emitted scalar concentration field $c(\bm{x},t)$ is rendered in the background using a grey-shaded logarithmic color scale. Color intensity along the trajectory encodes elapsed time since departure, with darker gold tones indicating later decision steps. Small red squares mark individual detection events ($\tau_d = 0$), i.e.\ moments at which the agent registers a non-zero odor signal above threshold. Time is reported above the panel in units of the agent's decision time step. This trajectory corresponds to the episode shown in Fig.~3(c) of the main text.\\

\textbf{Supplementary Movie~S2 $|$ Representative trajectory of the Q-RL agent in the mild mean wind regime ($U/u_\mathrm{rms} = 1$, $\tau_w = 100$).} Same as \emph{Supplementary Movie~S1}, but for an agent operating at $\tau_w = 100$ (teal trajectory).\\

\textbf{Supplementary Movie~S3 $|$ Representative trajectory of the cast-and-surge agent in the mild mean wind regime ($U/u_\mathrm{rms} = 1$, $\tau_w = 30$).} Same as Supplementary Movie~S1, but for the cast-and-surge heuristic with optimal casting angle ($\theta_c=70^{\circ}$) and wind-estimate memory time $\tau_w = 30$ (blue trajectory), run on the same episode; the agent fails to locate the source within the time horizon.\\

\textbf{Supplementary Movie~S4 $|$ Representative trajectory of the Q-RL agent in the isotropic regime ($U/u_\mathrm{rms} = 0$, $\tau_w = 5$).} The agent (yellow-red trajectory) starts at the red square and navigates toward the odor source (green circle), whose emitted scalar concentration field $c(\bm{x},t)$ is rendered in the background using a grey-shaded logarithmic color scale.
Color intensity along the trajectory encodes elapsed time since departure, with darker red tones indicating later decision steps. Small red squares mark individual detection events. Time is reported above the panel in units of the agent's decision time step. This trajectory corresponds to the episode shown in Fig.~4(d) of the main text.\\

\textbf{Supplementary Movie~S5 $|$ Representative trajectory of the Q-RL agent in the isotropic regime ($U/u_\mathrm{rms} = 0$, $\tau_w = 30$).} Same as \emph{Supplementary Movie~S4}, but for an agent operating at $\tau_w = 30$ (yellow-green trajectory).\\

\textbf{Supplementary Movie~S6 $|$ Representative trajectory of a spiraling agent in the isotropic regime ($U/u_\mathrm{rms} = 0$).} Same as Supplementary Movie~S4, but for the spiral-search heuristic (blue trajectory) with optimal growth rate ($\delta\ell=8$).\\

\textbf{Supplementary Movie~S7 $|$ Time evolution of the passive scalar field in the mild mean wind regime ($U/u_\mathrm{rms} = 1$).}
The scalar concentration field $c(\bm{x}, t)$ is shown in log-scale at the full DNS resolution ($1024^2$ grid), with the source located at the yellow star; the mean wind blows from right to left.
The plume is continuously emitted from the point-like source and stretched and folded by the underlying turbulent velocity field into an irregular, intermittent filamentary structure, illustrating the sparse and unpredictable nature of the odor signal encountered by navigating agents. The absorbing penalization mask applied near the domain boundaries, which prevents scalar re-entrance through the periodic boundaries, is visible as the sharp drop in concentration at the edges. Time $t$ is indicated in simulation units above the panel.\\

\textbf{Supplementary Movie~S8 $|$ Time evolution of the passive scalar field in the isotropic regime ($U/u_\mathrm{rms} = 0$).} Same as Supplementary Movie~S7, but in the absence of a mean wind.

\end{document}